\documentclass{aa}
\usepackage{graphicx} 
 \def\4u{4U~1543--47}
 
 \def\3c8{3C~88}
 \def\3c4{3C~444}
 
 \def\oiii{[{\sc O\, iii}]}
 \def\oii{[{\sc O\, ii}]}
 
 \def\feka{Fe K$\alpha$}

 \def\rxte{{\it RXTE}}

 \def\oi{\relax \ifmmode {\rm O\,{\sc i}}\else O\,{\sc i}\fi}
 \def\oii{\relax \ifmmode {\rm O\,{\sc ii}}\else O\,{\sc ii}\fi} 
 \def\oiii{\relax \ifmmode {\rm O\,{\sc iii}}\else O\,{\sc iii}\fi} 
 \def\ltsima{$\; \buildrel < \over \sim \;$}
 \def\simlt{\lower.5ex\hbox{\ltsima}} 
 \def\gtsima{$\; \buildrel > \over \sim \;$}
 \def\simgt{\lower.5ex\hbox{\gtsima}} 

\begin{document}

\title{Characterizing black hole variability with nonlinear methods:\\
the case of the X-ray Nova 4U~1543--47}

\author{M. Gliozzi\inst{1} 
\and C. R\"ath\inst{2}  
\and  I.E. Papadakis\inst{3,4}
\and P. Reig\inst{4,3}} 
\offprints{mario@physics.gmu.edu} 
\institute{George Mason University, 4400 University Drive, Fairfax, VA 22030
\and Max-Planck-Institut f\"ur extraterrestrische Physik, Postfach 1312, 
D-85741 Garching, Germany 
\and Physics Department, University of Crete, 710 03 Heraklion,
Crete, Greece
\and
Foundation for Research and Technology - Hellas, 
IESL, Voutes, 71110 Heraklion, Crete, Greece}

\date{Received: ; accepted: }

\abstract 
{}
{We investigate the possible nonlinear variability properties of the 
black hole
X-ray nova 4U~1543-47 with a dual goal: 1) to complement the 
temporal studies based on linear techniques, and 2) to search for signs of 
(deterministic and stochastic) nonlinearity in Galactic black hole 
(GBH) light curves. The proposed analysis may provide
additional model-independent constraints to shed light on black hole systems
and may strengthen the unification between  GBHs
and Active Galactic Nuclei (AGN).}
{First, we apply the weighted scaling index method (WSIM) to characterize the
X-ray variability properties of \4u\ in different spectral states
during the 2002 outburst. Second, we use 
surrogate data to investigate whether the variability is nonlinear in 
any of the different spectral states.}
{The main findings from our nonlinear analysis can be summarized as follows: 
1) The mean weighted scaling index $\langle\alpha\rangle$ appears to be 
able to parametrize uniquely the temporal variability properties of 
GBHs. The 3 different spectral states of the 2002 outburst of \4u\
are characterized by different and well constrained values of
$\langle\alpha\rangle$ satisfying the following relationship:
$\langle\alpha\rangle_{\rm VHS}~ < ~\langle\alpha\rangle_{\rm LS}~ <
\langle\alpha\rangle_{\rm HS}$. 
2) The search for nonlinearity  reveals that
the variability is linear in all light curves with the notable exception
of the very high state (VHS).
} 
{Our results imply that we can use the WSIM to assign
a single number, namely the mean weighted scaling index 
$\langle\alpha\rangle$, to a light curve,
and in this way discriminate among the different spectral states of a source.
The detection of nonlinearity in the VHS, that is characterized by the presence
of most prominent QPOs, suggests that intrinsically linear models which have been
proposed to account for the low frequency QPOs in GBHs are probably ruled out. 
Finally, the fact that WSIM results are scarcely affected by the noise level and
length of the light curve, naturally suggests an application to AGN variability
with the possibility of a direct comparison with GBHs. However, before deriving
more general conclusions, it is first necessary to carry out a systematic
nonlinear analysis on several GBHs in different spectral states in order to
assess whether the results obtained for 4U 1543-47 can be considered as
representative for the entire class of GBHs.}

\titlerunning{Nonlinear variability study of \4u}
\authorrunning{M.~Gliozzi et al.}
\maketitle

\section{Introduction}
In recent years, several studies have demonstrated the importance of X-ray 
temporal and spectral studies of black hole systems.  Because of their 
closeness and brightness, the physical conditions of Galactic black holes 
(GBHs) are better known 
than those of supermassive black holes in active galactic nuclei
(AGN), and in principle can be used to infer information about 
their scaled-up extragalactic analogs. For example, is it now well 
accepted
that GBHs undergo a continuous spectral evolution, switching between two 
main states: the ``low/hard" (hereafter LS) and the ``high/soft" (HS) passing 
through less well-defined
and short-lived ``intermediate states'', sometimes called steep power law, 
SPL, 
state or very high state, VHS, if occurring at high luminosity
(see  McClintock \& Remillard 2006, and Done et al. 2007 for recent 
comprehensive reviews on GBHs). 

Despite a substantial advance in this field, however, several questions
yet remain unanswered. For example, it is currently strongly debated 
whether the LS is
characterized by a truncated disk or not (e.g., Miller et al. 2006a,b;
Gierli\'nski et al. 2008; D'Angelo et al. 2008).
It is still unknown if the jet plays an important role
in the X-ray range during the LS (e.g., Markoff et al. 2001;  
Zdziarski et al. 2003), and whether there is a physical difference 
between the LS and the quiescent state (e.g., Tomsick et al. 2004;
Corbel et al. 2006). It is also still unclear 
what is the origin of QPOs, which only appear in specific spectral states.
Finally, we still have a poor knowledge of the physical conditions of the 
accretion flow in the VHS,  which is often 
associated with the most powerful relativistic ejections
(e.g., Fender et al. 2004). 

Several different models, mostly driven by  X-ray spectral results,
have been proposed to explain the aforementioned
open questions. However, due to the transient nature and short duration of
these phenomena, the spectral information alone is unable to discriminate
between competing models, leading to the so-called ``spectral degeneracy''.
 
In order to break this degeneracy, it is therefore of crucial importance to 
complement the spectral information with additional constraints from 
the temporal analysis. In this framework, the use of the power spectral 
density (PSD) functions has proved to be very successful:
the combined  temporal and spectral information has led to a generally accepted
scenario where the spectral evolution of GBHs is mostly driven by variations
of the accretion rate $\dot m$, which in turn lead to changes in 
the interplay between accretion disk, Comptonizing corona, and a relativistic
jet (e.g.,  McClintock \& Remillard 2006).

The success of the PSD in GBH and AGN studies (see, e.g., 
Klein-Wolt \& van der Klis 2008; McHardy et al. 2006), emphasizes the
importance temporal studies and the crucial role that they may play
in breaking the spectral degeneracy. It is however important to 
explore also alternative temporal methods since the PSD, as any
other timing techniques, cannot exhaustively characterize
any non-trivial variable system. For example, since
the PSD is sensitive only to the first 2 moments of the probability 
distribution, it cannot provide a complete description of non-Gaussian
processes. Similarly, since the PSD is an intrinsically linear technique
it is not adequate to characterize systems with non-linear variability.

The main goal of this work is to investigate alternative temporal methods,
that are complementary to the PSD.
More specifically,
in the first part of this paper ($\S3$, and $\S4$), we carry out a 
nonlinear analysis of the variability properties of the black hole X-ray nova 
\4u\ (whose data description is provided in $\S2$) by using the weighted 
scaling index method (WSIM). 
The scaling index method (e.g., Atmanspacher et al. 1989) has been 
employed in a number of different fields because of its ability to discern
underlying structure in noisy data. For example, it has been successfully
used in medical science (see Morfill \& Bunk 2001 for a brief review), in 
image analysis (e.g., R\"ath \& Morfill 1997; Brinkmann et al. 1999), 
in plasma physics (e.g., Ivlev et al. 2008; S\"utterlin et al. 2009),
and in
cosmology (R\"ath et al. 2002, 2007, 2009). Recently, we have applied this method to 
well-sampled AGN light curves to look for signs of nonstationarity 
(Gliozzi et al. 2002; 2006).

In this work, our primary goal is to 
investigate whether we can parametrize the variability properties 
of a GBH with a single number, similarly to what is currently done in the
spectral analysis where the energy spectra of accreting objects are 
usually characterized by one number, namely the slope of the power-law 
model. To this aim, we apply  the WSIM to \4u, a BH system
for which the correspondence between spectral states and individual 
observations  is well defined during its outburst in June 2002.
If the results from this analysis are encouraging, we plan
to carry out a similar systematic analysis on a sample of BH systems, to
investigate whether a common pattern emerges.

The second part of this works ($\S5$)
deals with the search for nonlinearity in
the light curves of 4U 1543-47, in all its spectral states. We use a new
method to produce reliable surrogate data and the nonlinear prediction
error (NLPE; Sugihara \& May
1990) test to search for any signs of non-linearity. As discussed
below, this statistic is able to detect any non-linear behavior in the
light curves, irrespective of whether the non-linearity is ``deterministic" or
``stochastic" in nature. 

It is worth noting that, unlike previous nonlinear studies of GBHs (e.g,
Misra et al. 2004; 2006), our goal is not to search for low-dimensional chaotic
signatures in GBHs, but to characterize the global variability properties of 
BHs in a simple and scalable way.
We stress that this kind of analysis is not in contrast with
the ``standard'' linear analysis, whose contribution is of crucial
importance in guiding our work,
but rather it complements it by exploring aspects of 
the variability that are not accessible to linear techniques.

\section{Data Description}
\4u\ is a recurrent X-ray nova, with outbursts occurring every 10--12 years.
Dynamical optical studies during quiescence yield a primary mass of $M_1=
9.4\pm2.0 M_\odot$ strongly arguing for the presence of a BH 
(Park et al. 2004). 

For our analysis we will use \rxte\ PCA data of \4u\ during the outburst
occurred between June 17 and July 22, 2002, which corresponds to the interval
52,442--52,477 in Modified Julian Date (MJD= Julian Date-2,400,000.5). 
The \rxte\ PCA covered the 2002
flare of \4u\ with at least one observation per day with exposures
ranging between $\sim$800 s and $\sim$3900 s. More than 90\% of the 
observations caught the source in  high state (HS), whereas only a couple
of observations cover the short-lived very high state (VHS) and the 
beginning of the low state (LS).

A detailed analysis of the energy and power spectral densities was performed
by Park et al. (2004).  For our purpose, the relevant results of their 
work can be summarized as follows: 1) The source was caught by the PCA
close to the outburst peak, when \4u\ was in a thermally dominated state
HS. 2) For nearly the entire duration of the outburst, the source remained 
in the HS, which is temporally characterized by a featureless PSD
(see Figs. 8a,c of Park et al. 2004). 
The only notable exceptions are:  a rapid transition to the VHS around 
52,459--52,460 MJD, whose PSD shows a very prominent QPO around 
 5--10 Hz (Fig. 8b of Park et al. 2004)
, and  
a transition to the LS toward the end of the observation
showing the typical band-limited noise PSD
(Fig. 8d of Park et al. 2004). 3) The energy spectral analysis indicates 
that an acceptable parametrization of all spectra always requires a disk,
a power-law component, and a \feka\ line, 
whose relative contributions vary with
time. Specifically, the disk flux closely follows the total count
rate evolution 
during the outburst, whereas the flux associated with the power-law component
shows a broad and prominent peak during the VHS, preceded by  2 less 
prominent peaks. 

All light curves were extracted in the 2--20 keV energy range following the
standard \rxte\ procedure and were binned at a resolution of 
0.100097656 s (for simplicity hereafter we will use 0.1 s when referring to
the time bin), which is 205 times the original integration time. 
A more detailed description of the data reduction is
provided by Reig et al. (2006).

\section{Weighted Scaling Index Method}

Since a detailed description of the 
scaling index method (SIM) has already been provided in our previous work, 
here we limit ourselves to summarizing the main steps in a simple and
qualitative way and
pointing out the main differences introduced by the weighted scaling
index method (WSIM) that we use in this work.
\begin{figure}
\begin{center}
\includegraphics[bb=200 230 445 615,clip=,angle=0,width=8.cm]{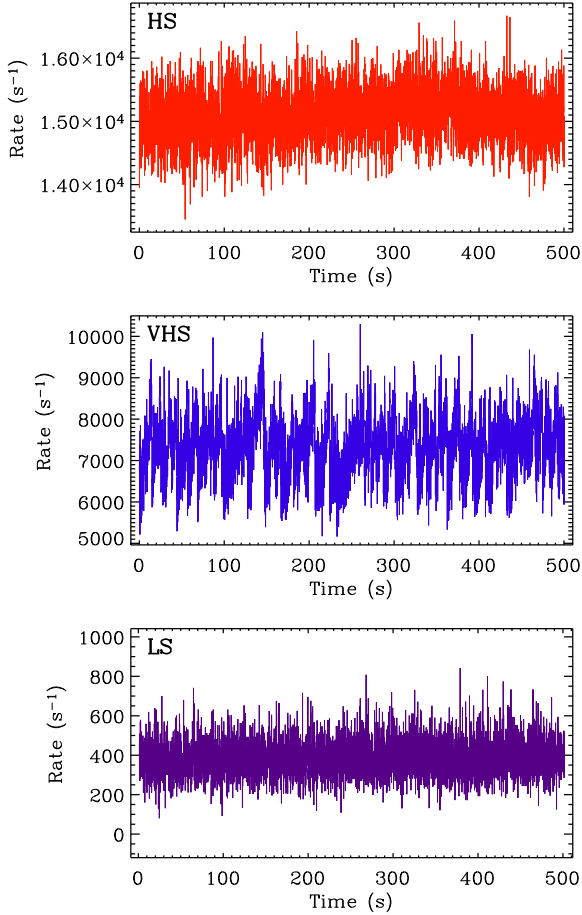}
\caption{Representative light curves during  HS (top panel), VHS (middle 
panel), and LS (bottom panel). Time bin is 0.1 s 
}
\end{center}
\label{fig1}
\end{figure}
\begin{figure}
\begin{center}
\includegraphics[bb=16 16 300 666,clip=,angle=0,width=6.cm]{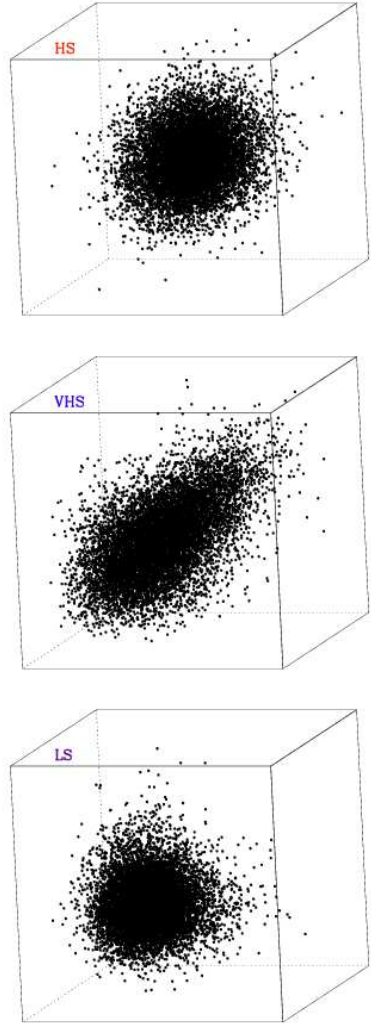}
\caption{3-dimensional phase space portraits for  HS (top panel), VHS (middle 
panel), and LS (bottom panel). $\tau=$0.1 s was used for the time-delay
reconstruction.
}
\end{center}
\label{figure:fig2}
\end{figure}

\begin{enumerate}

\item As all timing techniques, the very first step starts with a time
series. Figure~1 
shows three samples of light curves
characterizing HS (top panel), VHS (middle panel), and LS (bottom panel)
of \4u\ during the 2002 outburst. For clarity reasons, 500 s time intervals
(i.e., intervals with 5000 data points) are plotted, although the results
described in $\S$4.1 are obtained using 1000 s (10,000 points) intervals;
the time bin used for all light curves is 0.1 s. Before
applying the SIM, the light curves are normalized, in the sense that the
mean count rate is subtracted from each data point and the resulting quantity
is divided by the total standard deviation. In any variability analysis,
particular attention should be paid to low count rate intervals due to the 
Poisson noise that may swamp the signal and hamper the analysis.
Despite the large difference in count rate and hence in Poisson noise
(during the 2002 outburst the average \rxte\ PCA count rate during individual
observations ranges between $\sim$18,000--400 count/s), the  SIM analysis
appears to be largely unaffected by the actual count rate (see $\S$4.2
for more details).

\item As most of the methods of nonlinear dynamics applied to timing analysis,
the SIM relies upon the phase space reconstruction, which is obtained via the 
time-delay reconstruction (e.g., Kantz, \& Schreiber 1997; Regev 2006). 
In 
simple words, in the case of a 3-dimensional 
(3-D)
phase space reconstruction, one constructs a set of 3-D vectors by selecting
triplets of data points from the original time series, which are separated 
in time by the time-delay $\tau$. More specifically, the second data-point
(which represents the y-component in a 3-D vector) is separated from 
the first one (the x-component) by a time delay of $\tau$, whereas the third 
data-point
(the z-component) is separated from the first one by 2 $\tau$, and from the
second one by $\tau$. Figure~2 illustrates the phase space
portraits in 3 different spectral states, i.e. the results of a 3-D phase space
reconstruction for the light curves shown in  Figure~1 using a
time delay of $\tau=0.1$ s. Not surprisingly, just like the light curve
plotted in the middle panel of Fig.~1 looks different from the other two,
the phase space portrait of
the VHS looks considerably different from those describing HS and LS. This
reflects the fact that, unlike HS and LS, the VHS light curve is characterized
by the presence of frequent peaks and dips in a quasi-periodic fashion, 
which increase the degree of correlation observed in the 
phase space reconstruction.

\item After the phase space reconstruction, the temporal properties of the
original time series translate into topological properties of the phase space
portraits. In order to quantify these topological properties, a useful method 
is based on the measure of the crowding  around each individual 
vector. This measurement is formally performed by computing the cumulative 
number 
function, $C_i(R)=n\{j\vert d_{ij}\leq R\}$, which measures the number of 
vectors $j$, whose distance $d_{ij}$ from a vector $i$ is smaller than $R$.
Generally, when plotted versus $R$ in log-log space,
the function $C_i(R)$ can be approximated by a power law, 
$C_i(R)\propto R^{\alpha_i}$, for a wide range of values of $R$. The exponent,
$\alpha_i=[\log(C_i(R_2))-\log(C_i(R_1))]/[\log(R_2)-\log(R_1)]$, which is the 
logarithmic derivative of the cumulative number 
function, is the scaling index.
In summary, for a light curve with $N$ data points and an embedding space
of dimension $D$ this process yields $N-D$ values of $\alpha_i$. The temporal
properties of the original light curve can then be studied either using the
distribution of $\alpha_i$ or the mean value of this distribution 
$\langle\alpha\rangle$, as explained in $\S$4.

\item The main difference between  ``normal'' and weighted scaling 
index methods is that in the latter the cumulative number function is 
substituted by the weighted cumulative point distribution, where the
weighting function can be any differentiable function (in our case a
Gaussian function; see R\"ath et al. 2007 for a detailed explanation
of the WSIM). The chief advantage of WSIM is twofold: 1) the logarithmic 
derivative (i.e., the scaling index) can be computed analytically instead
of numerically; 2) the number of free parameters is reduced by 1: since the
logarithmic derivative is computed analytically, we only need to define
one value $R$ at which it is computed, instead of $R_1$ and $R_2$.
\end{enumerate}

To summarize, i) we start from normalized light curves, ii) transform them into
phase space portraits via time-delay reconstruction, and iii) quantify their
differences by computing the weighted scaling index.

Importantly, the mean value $\langle\alpha\rangle$  provides direct
information on the nature of the variability process: for a purely random 
process $\langle\alpha\rangle$ tends to the value of the dimension of the 
embedding space (i.e., the space used in the phase space reconstruction)
, whereas for correlated (and deterministic) processes 
$\langle\alpha\rangle$ is always smaller 
than the dimension, $D$, of the embedding space and, in the ideal case where
the random component is completely negligible, the mean scaling index is
independent of the embedding dimension.
In other words, low values of $\langle\alpha\rangle$ characterize correlated
variability processes, whereas
higher values correspond to 
variability properties with a higher degree of ``randomness''.

Before discussing the results of the WSIM, it is important to understand
the role played by
the 3 free parameters involved in this process and their impact on the results.
The process leading to the phase space reconstruction requires to choose
2 parameters,
the time delay $\tau$ and the embedding dimension $D$, and the computation
of the weighted scaling index requires an additional parameter, the radius
$R$ at which the logarithmic derivative is computed. 

A common choice for 
$\tau$ is the characteristic timescale of the system, which can be determined
with different methods (e.g., autocorrelation function, PSD, or the so-called
mutual information; Fraser \& Swinney 1986). In our analysis we will use
$\tau=0.1$ s, which is where the PSDs of \4u\ show prominent QPOs in the
VHS and LS, whereas as expected the PSD of the HS is featureless
(see Park et al. 2004). 
On the other hand, there are no
systematic prescriptions for the choice of $D$ and $R$. The latter likely 
depends on the typical distances (in the following we will use the Euclidean 
norm as measure of the distance)
between vectors, which, in turn, depend on 
the choice of the embedding space dimension. 

In principle, the discriminating power of the SIM is enhanced 
by using high embedding dimensions. However, for
our purposes, a relatively low embedding dimension is preferable,
since we work with a limited number of points and since one of our  goals is 
to compare the GBHs variability results with those of AGN, which have
light curves characterized by fewer data points than GBHs. Specifically, for
our analysis we utilize $D=3$. 
Once $\tau$ and $D$ are fixed, $R$ is obtained in
the following way: first, the temporal order of the data points in the 
original time series is randomized creating 10 sets of randomized data;
second, the mean WSI is computed for randomized and original data using
different values of $R$ (specifically between 0.5 and 2.5); finally, the
chosen radius (in our case $R=1.6$) is the
value that yields the larger difference between randomized data and
original data, indicating that it is the value most sensitive to the temporal
structure of the original data. 

Our nonlinear analysis of the variability properties of \4u\ is therefore 
carried
out using $\tau=0.1$ s, $D=3$, and  $R=1.6$. However, for the sake of
completeness, we have performed an investigation of a broad parameter
space encompassing $\tau=0.1-10$ s, $D=2-4$, $R=0.6-2.4$. 
The results of this 
analysis, which demonstrates that our main findings are largely insensitive
to the choice of these 3 parameters, are reported in the Appendix.

\begin{figure}
\begin{center}
\includegraphics[bb=200 100 445 740,clip=,angle=0,width=6.cm]{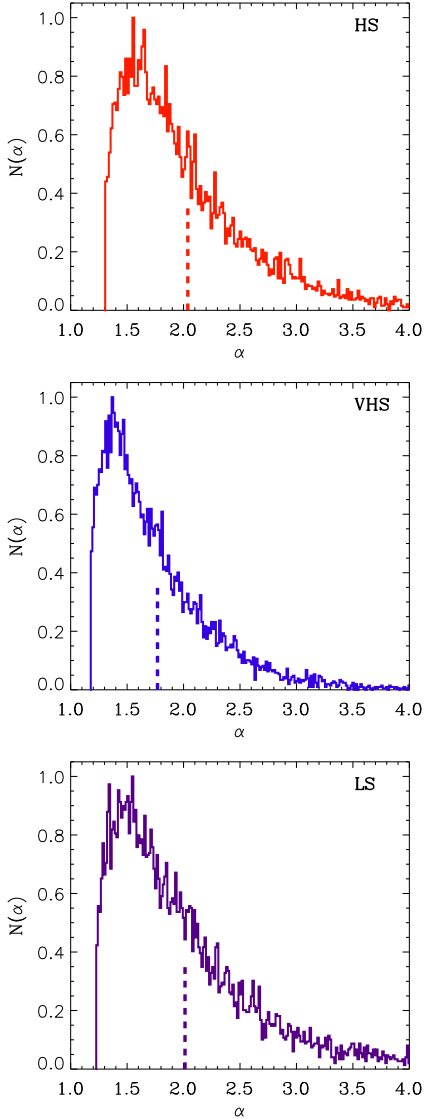}
\caption{Histograms of the weighted scaling indices (WSIs) for  
HS (top panel), VHS (middle panel), and LS (bottom panel). The dashed lines
represent the mean values. All WSI values were computed for a 3-D
embedding space, $\tau=$0.1, and $R=1.6$, using light curves containing
10,000 data points.  
}
\end{center}
\label{fig3}
\end{figure}
\section{WSIM Results}
As explained before, the temporal properties of a
variable system can be studied either via the distribution of the $\alpha_i$ 
values or simply through the mean value $\langle\alpha\rangle$. 
In the following we
elucidate these 2 approaches, by applying the WSIM first
to the 3 representative light curves shown in Figure~1, and then to all
light curves covering the 2002 outburst of \4u.

\subsection{WSIM distribution}
Given that each of the representative
light curve has 10,000 points
each and given that the WSIM is applied to each
individual vector, this analysis yields nearly 10,000 values of $\alpha_i$.
The results of this process are illustrated in
Figure~3 which shows the $\alpha_i$ distributions for the HS (top panel),
VHS (middle panel), and LS (bottom panel); the dashed lines represent the 
mean value $\langle\alpha\rangle$ in the 3 different spectral states.

At first order, all 3 histograms share a similar asymmetric shape with a 
sharp cut-off on the left-hand side and a broad right-hand tail. The 
left-hand side of the $\alpha_i$ distribution is generally related to the 
correlated variability component, whereas the right-hand side is related
to the random noise component. As a consequence, a highly correlated
process is characterized by a narrow $\alpha_i$ distribution peaking at
low values. On the other hand, a process dominated by random noise will
be associated to a broad $\alpha_i$ histogram with a pronounced right-hand
tail extending to large values of $\alpha$.

A first simple way to assess the difference between the 3 
representative histograms is to compare their respective means. For
the HS, VHS, and LS we get respectively  $\langle\alpha\rangle= 2.038\pm0.006,
~1.769\pm0.005,~2.013\pm0.007$ (where the quoted uncertainties are
$\sigma/\sqrt{n}$). These values suggest that the VHS is significantly 
different from both the HS (at $\sim 33\sigma$ level) and the LS
(28 $\sigma$ level), whereas the difference between HS and LS is only
marginally significant according to this test (2.7$\sigma$).

A formal comparison between the 3 distributions 
based on a Kolmogorov-Smirnov test (hereafter K-S test), which 
is sensitive to the whole distributions of $\alpha_i$
suggests that the three distributions
are statistically different from each other. Specifically, the K-S test
yields a statistic of 0.22
and an associated probability $P_{\rm K-S}< 10^{-25}$ that 
HS and VHS $\alpha_i$ histograms are drawn from the same distribution. 
Similarly, for HS vs. LS and VHS vs. LS we obtain 0.07 
($P_{\rm K-S}\simeq10^{-24}$)
and  0.16 ($P_{\rm K-S}< 10^{-25}$), respectively.
However, it must be kept in mind that the K-S test is devised for 
independent
data-points, whereas the different $\alpha_i$ are not completely independent.
As a consequence, the apparently highly significant difference between 
the 3 histograms representing 3 different spectral states should be 
considered with caution and need to be confirmed with further analysis
(see below).

The results from the histogram analysis are encouraging and suggest that the 
WSIM has indeed the potential to discriminate 
between the different spectral states of \4u, in full 
agreement with results from the PSD analysis. However, in order to reach
a stronger conclusion, we should demonstrate that all $\alpha_i$  
histograms of HS are
indistinguishable from each other, but statistically different from all
the VHS and LS  $\alpha_i$ histograms. Although feasible, this procedure
would be very time consuming and would go against the primary goal
of this work, which is to provide a {\it simple} alternative  way
to characterize the temporal properties of GBHs. In addition,
it must be pointed out that these results have been obtained using 
10,000-point light curves, which are generally not commensurable with 
typical AGN light curves. This approach would therefore hamper a direct 
comparison between GBH and AGN,
which is one of the secondary goals of this work.

\subsection{WSIM mean}
Since our primary goal is to define a {\it simple} way to characterize the 
global
variability properties of GBHs and since in this kind of analysis
the mean value $\langle\alpha\rangle$ is the most robust 
indicator of the global variability properties, we will restrict our analysis 
to $\langle\alpha\rangle$. In this way, the 
properties of a given light curve are defined by a single number, 
$\langle\alpha\rangle$, in a
similar way as the photon index is often used to characterize the energy
spectral properties of X-ray sources. 

\begin{figure}
\includegraphics[bb=100 25 460 400,clip=,angle=0,width=9.cm]{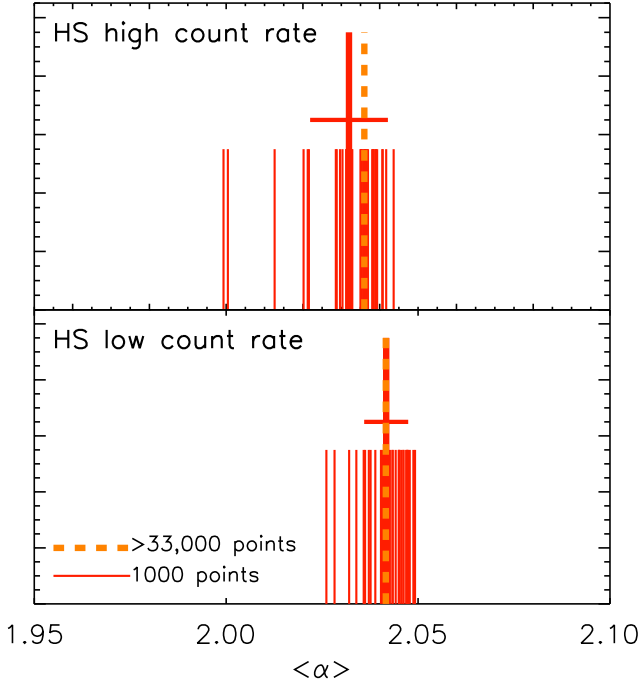}
\caption{Comparison between the mean scaling index obtained using long
light curves ($>$ 33,000 points; thick pale-colored dashed lines) and the 
values derived using intervals with
1000 points (smaller dark solid lines), whose average is represented by the 
thick
solid lines with horizontal lines indicating the dispersion $\sigma$ . 
The top panel refers to a high count rate HS ($\sim$16,000 counts/s), 
whereas the bottom panel 
describes results from a low count rate HS ($\sim$800 counts/s). All values of 
$\langle\alpha\rangle$
were computed for a 3-D embedding space, $\tau=$0.1, and $R=1.6$
}
\label{figure:fig4}
\end{figure}
\subsubsection{Test with short intervals}
In addition, in order to further generalize this procedure extending it
to relatively short light curves, which are more common than long 
uninterrupted ones, we will use 
intervals of 100 s (i.e., time series containing 1000 points since the bin 
time is 0.1 s). In this way, the light curves will contain a number of points 
comparable to typical AGN light curves, offering the possibility of a direct 
comparison between GBH and AGN variability properties

Before proceeding further, we must first demonstrate that the choice of shorter
intervals will not hamper our analysis. On one hand, from the PSD analysis
of Park et al. (2004) we are ensured that nothing relevant occurs to the
timing properties of \4u\ for timescales longer than $\sim$ 100 s: all the 
interesting
PSD features (frequency breaks and QPOs) are located at frequencies well
above $10^{-2}$ Hz. On the other hand, we must still verify that the WSIM 
results
are not significantly affected by the use of shorter intervals. 

\begin{figure}
\includegraphics[bb=50 60 460 430,clip=,angle=0,width=8.cm]{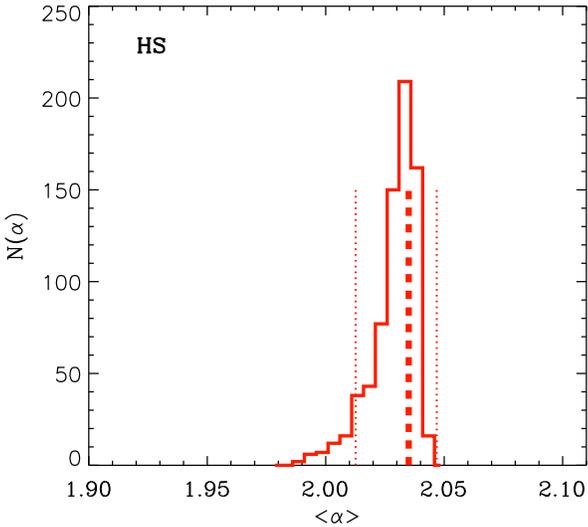}
\caption{Distribution of the mean WSI values, obtained using segments
of 100 s (1000 points) during  HS (741 segments). All values of 
$\langle\alpha\rangle$ were computed for a 3-D embedding space, 
$\tau=$0.1, and $R=1.6$.
The thick dashed line indicates the mean of the $\langle\alpha\rangle$
distribution and the dotted lines represent the 90th percentiles.
}
\label{figure:fig5}
\end{figure}
To this aim,
and to demonstrate that the WSIM is also independent of
the mean count rate (and hence of the Poisson noise level),
we have performed the following experiment. We have chosen 2 of the
longest HS light curves (both have more than 33,000 data points) with
mean count rate greatly different: the first light curve, obtained close
to the outburst peak, has an average \rxte\ PCA count rate of $\sim$16,000
counts/s, whereas the second one (corresponding to a later phase of the decay)
has a count rate of only $\sim$800 counts/s.
We have applied the WSIM to both data sets, first using the entire light
curve and then using intervals of 1000 points only. The results are 
illustrated in Figure~\ref{figure:fig4}, where the value of 
$\langle\alpha\rangle_{33,000}$ for
the entire light curve is represented by the thick dashed line, the 
individual values  obtained with 1000 points, $\langle\alpha\rangle_{1000}$, 
are depicted as shorter continuous lines, and their average is indicated by the
thick continuous line. The horizontal lines indicate the standard deviation,
$\sigma_{\langle\alpha\rangle_{1000}}$, of the sample of
$\langle\alpha\rangle_{1000}$. Figure~\ref{figure:fig4}
reveals that:\\
\indent 1) In both cases, $\langle\alpha\rangle_{1000}$ values narrowly 
cluster around the mean scaling index, $\langle\alpha\rangle_{33,000}$, 
obtained from the entire light curve and 
their average,
$\langle\langle\alpha\rangle_{1000}\rangle$, is fully consistent with 
$\langle\alpha\rangle_{33,000}$. This is formally demonstrated
by the fact that
$\vert\langle\alpha\rangle_{33,000}-\langle\langle\alpha\rangle_{1000}\rangle\vert/(\sigma_{\langle\alpha\rangle_{1000}}/\sqrt{n})=2.5$ (for the high count rate case)
and 0.1  (for the low count rate case), which are both lower 
than the 3$\sigma$ level. This indicates that  WSIM results obtained with
100 s intervals are fully consistent with those obtained using a longer 
interval.
Therefore, with this method the variability properties
of \4u\ can be thoroughly investigated using 100 s intervals.\\ 
\indent 2) Although visually the high and low count rate distributions
of $\langle\alpha\rangle_{1000}$ and their respective mean
look fairly close and indeed their standard deviations significantly overlap,
statistically speaking, their difference 
$\vert\langle\langle\alpha\rangle_{\rm 1000,high}\rangle-\langle\langle\alpha\rangle_{\rm 1000,low}\rangle\vert$ is slightly above the formal
3$\sigma$ level. 
In order to 
thoroughly address this issue, and estimate quantitatively the uncertainty 
on the scaling
index during the HS, we need to account for {\it all} the observations during
HS. This is illustrated in Figure~\ref{figure:fig5}, which shows the 
distribution of $\langle\alpha\rangle_{1000}$ obtained using
all the 100 s intervals
of all the available HS observations. Despite the huge difference in count
rate (max$>$ 18,000 counts/s, min$<$370 counts/s) and a temporal separation
longer than 30 days,
the vast majority of values 
narrowly clusters around the mean, yielding
$\langle\langle\alpha\rangle_{\rm 1000}\rangle=2.035_{-0.022}^{+0.012}$,
where the quoted uncertainties are the 90th percentiles
(the error on the mean is $\sigma/\sqrt{n}=0.01/\sqrt{741}=0.0004$).
The ``narrowness" of the 
distribution of the $\langle\alpha\rangle$ values shown in Fig.~5, 
really suggests that WSIM
results are not affected neither by the fact that the light curves have vastly
different signal-to-noise ratios nor by the time span over which the HS light
curves are spaced. It is therefore suggestive that 
$\langle\alpha\rangle$ is most probably
determined just by one factor, i.e. the properties of the variability mechanism
during the HS.

For completeness, we also carried the previous test for the two VHS and LS
light curves.  Note that the short duration of the VHS (due to the intrinsic 
transient behavior of this short-lived state) and the LS (due to the 
interruption in the PCA monitoring program) yielded only 2 light curves
for each state.
In both cases,  $\langle\langle\alpha\rangle_{1000}\rangle\rangle$ is fully 
consistent with 
$\langle\alpha\rangle_{\rm whole}$, as their difference is of the order
of 1 $\sigma$ or less.  Similarly, the values of the difference between
the pair of light curves corresponding to the same state,
$\vert\langle\langle\alpha\rangle_{\rm 1000,A}\rangle-\langle\langle\alpha\rangle_{\rm 1000,B}\rangle\vert/\sqrt{\sigma_{\rm A}^2+\sigma_{\rm B}^2}$, is 
0.07 and 1.4 for the VHS and LS, respectively.

\subsubsection{Discrimination of spectral states}
We can now assess whether the WSIM is able to discriminate
between the spectral different states, by considering all the
available observations, dividing all the individual light curves covering 
the 2002 outburst into 100 s intervals and treating each segment as an 
independent data-set. This
procedure yields 741 data sets for HS, 31 for VHS, and 40 for LS. 

\begin{figure}
\includegraphics[bb=105 60 515 375,clip=,angle=0,width=8.cm]{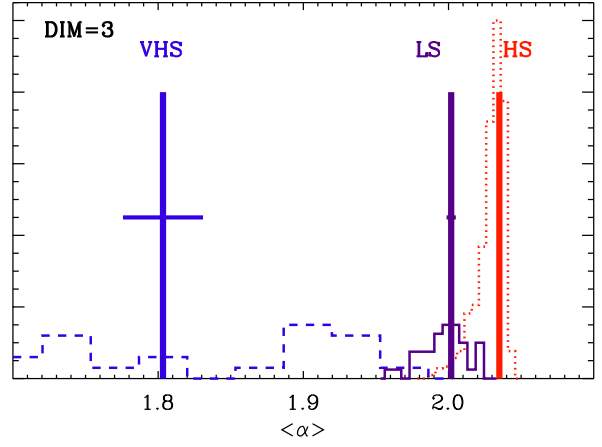}
\caption{Normalized distributions and averages of mean WSIs for HS, VHS, 
and LS, obtained using segments
of 100 s (1000 points) from all the available light curves. The uncertainties
shown represent the error on the mean (i.e., $\sigma/\sqrt{n}$, where $n$ is 
the 
number of data sets); for HS the uncertainty is smaller than the thickness
of the vertical bar representing the mean.
All values were computed for a 3-D embedding space, $\tau=$0.1, and $R=1.6$.
The maximum of the HS histogram has been normalized to 1, whereas 
the VHS and LS have been normalized to 0.15, only for clarity reasons.
No visual comparison would have been possible maintaining the real proportions
between the distributions, given that VHS and LS only contain 30--40 
data-points as opposed to the 741 contained in the HS}.
\label{figure:fig6}
\end{figure}

Figure~\ref{figure:fig6} shows the normalized distribution of 
$\langle\alpha\rangle_{1000}$ values for the HS (dotted line), LS (solid
line), and VHS (dashed line), as well as their average of the mean WSIs.
The HS and LS distributions are very narrow, and they appear to be
offset, with the LS values of $\langle\alpha\rangle$ being 
systematically lower than the respective values of HS.
The VHS distribution has a rather large width, but is this mainly due to
the fact that, in addition to the 2 VHS observations, we also included here
the data  from the  observation just before the source entered fully the VHS
state, which caught the source during a transition phase (see $\S$4.3). In any 
case though, the VHS values of $\langle\alpha\rangle$ appear to be 
systematically much lower than the values in the LS and/or HS.

A simple and robust way to quantify the difference between the 3 different
states is to compare their respective means by computing the quantity
$\Delta\alpha_{\rm A-B}\equiv\vert\langle\alpha_{\rm A}\rangle-\langle\alpha_{\rm B}\rangle\vert/\sqrt{\sigma_{\rm A}^2+\sigma_{\rm B}^2}$, where 
$\langle\alpha_{\rm A}\rangle$ is the mean scaling index obtained using all
the 100 s intervals during the spectral state $A$ , and $\sigma^2_{\rm A}$ is 
the variance divided by the number of 100 s intervals in that state. 
This test yields respectively $\Delta\alpha_{\rm VHS-HS}=8.4$,
$\Delta\alpha_{\rm VHS-LS}=7.2$, and $\Delta\alpha_{\rm HS-LS}=10.3$,
indicating that the WSIM is indeed able to statistically discriminate between
the 3 states.

The significance of the difference between the WSIs in
the three spectral states can also be examined using a  
K-S test, which can be safely applied to the 3 distributions of 
$\langle\alpha_i\rangle$,
since each data-point has been obtained from a separate 100 s interval,
which in many cases are separated by a few days intervals,
and hence can be considered as being independent ``measurements''.
Specifically, the  K-S test
yields  0.93 ($P_{\rm KS}< 8\times10^{-24}$), 0.78 ($P_{\rm KS} <2\times10^{-10}$), and 0.75 ($P_{\rm KS}<6\times10^{-20}$), 
for the cases of VHS vs. HS, VHS vs. LS, and
VHS vs. LS. 

In summary, the main results of the analysis based on all available light
curves divided into 100 s intervals can be summarized as follows:\\
1) In all
3 spectral states $\langle\alpha\rangle < D$ (where
$D=3$ is the embedding space dimension). This result
implies the presence of correlated variability, which is expected given the 
red-noise trend observed in all PSDs.\\ 
2) The 3 spectral states have different mean scaling indices satisfying
the following relationship:
$\langle\alpha\rangle_{\rm VHS}~ < ~\langle\alpha\rangle_{\rm LS}~ <
\langle\alpha\rangle_{\rm HS}$. In addition to formally demonstrating 
that the scaling index method is able to discriminate between the spectral
states of \4u, this results 
indicates that the VHS is the state
characterized by the highest degree of correlated variability. Also this
result is somewhat expected from the linear temporal analysis,
given the presence of a  prominent QPO in the VHS
PSD. However, the low value of $\langle\alpha\rangle_{\rm VHS}$ might also
be related to non-linear correlations,  i.e., any temporal correlations 
that cannot be detected by the PSD or autocorrelation function and that 
manifest themselves as correlation in the Fourier phases.
This will be discussed in more detail in $\S5$.

\subsection{Temporal evolution of $\langle\alpha\rangle$}
\begin{figure}
\includegraphics[bb=75 55 445 655,clip=,angle=0,width=8.cm]{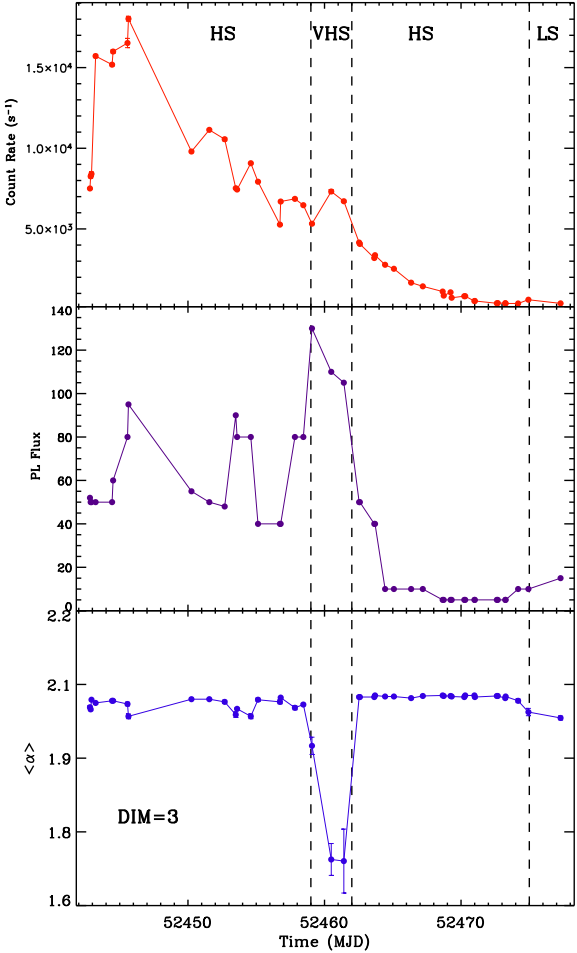}
\caption{Temporal evolution during the 2002 flare of \4u\ for the 
\rxte\ PCA count rate (top panel), flux associated with the PL component
in units of $10^{-10}~{\rm erg~cm^{-2}~s^{-1}}$
(middle panel), and mean WSIM (bottom panel). The statistical errors are in
most cases smaller than the symbols.
}
\label{figure:fig7}
\end{figure}
After having demonstrated that the WSIM is a reliable tool to characterize 
the variability properties in different spectral states, we focus now on the
temporal evolution of the mean scaling index during the 2002 flare of \4u.
 Since the source
spends the vast majority of the time in the thermal dominated HS,
no substantial changes
in the energy spectral parameters appear to occur during the outburst
(see Fig. 3 of Park et al. 2004). On one hand, this spectral behavior
represents a considerable advantage to thoroughly
assess the uncertainty on $\langle\alpha\rangle$, but on the other hand it
partially hampers a detailed
study of correlated spectral and temporal variability. Nevertheless, it is
interesting to investigate if (and how) the changes of 
$\langle\alpha\rangle$ are related to the flux changes during the
outburst evolution. The results of this analysis are illustrated in 
Figure~\ref{figure:fig7}:
the top panel describes the temporal evolution of the
total count rate that is 
roughly proportional to the evolution of 
disk flux and color temperature; the middle
panel shows the flux associated with the power law component
in units of $10^{-10}~{\rm erg~cm^{-2}~s^{-1}}$,
which was derived from Fig. 2 of
Park et al. (2004); finally, the bottom panel presents the 
$\langle\alpha\rangle$ temporal evolution. The scaling index analysis is
carried out dividing each individual observation into intervals of 100 s,
computing $\langle\alpha\rangle$ for each interval, and then taking the mean
over all intervals belonging to a given observation. The error-bars for the
mean scaling indices are given $\sigma/\sqrt{n}$ (where $n$ is 
the number of 100 s intervals of a given observation), which are often smaller
than the symbols shown in the bottom panel of Fig.~\ref{figure:fig7}.

The temporal trend of $\langle\alpha\rangle$ can be
summarized as follows: the mean scaling index remains roughly constant 
around 2.05 for the whole duration of the outburst, with the notable 
exception of a deep dip during the VHS state. A closer look at  
Fig.~\ref{figure:fig7} reveals 2 minor dips preceding the VHS and a steady
decrease toward the end of the outburst when the source is entering the LS.
Interestingly, while these changes of $\langle\alpha\rangle$ appear to be 
uncorrelated with the overall count rate (and hence with the disk flux), they
seem to occur in correspondence of local maxima in the power law flux. At
zeroth order, such apparent correlation  can be interpreted in the
following way: the variability associated with the power law component is 
``less random'' than the one produced by the disk. However, additional data
and a systematic analysis of correlated spectral and temporal properties is
necessary before drawing firmer conclusions.
 
Perhaps the most remarkable result from  Fig.~\ref{figure:fig7} is the fact 
that after $\sim$MJD 52462, while both the PL flux 
and the total count rate decrease very significantly (in an abrupt way
the PL flux and in a smoother way the disk flux), $\langle\alpha\rangle$
returns to the same level it was before the VHS state. In other 
words, $\langle\alpha\rangle$ appears to be a true indicator of state, 
as it is defined based on spectral and timing criteria.

\section{Search for Nonlinearity} 
The second goal of this work is to search for nonlinear temporal correlations  
in the variability properties of \4u.
Investigating the nature of the temporal variations is of primary interest
for constraining models of variability in GBHs (and AGN). Indeed, unlike
the results from the PSD and auto-correlation analyses, which can be 
equally well explained by a variety of different physical models, the 
detection of nonlinear variability would immediately rule out any 
intrinsically linear model that explain the X-ray variability as the
superposition of many independent active regions (e.g., Terrel 1972).
Therefore, this analysis has the potential to provide model-independent 
constraints that will break the current model degeneracy.

In order to find out whether a time series can be completely  
modeled by superimposed linear processes (plus uncorrelated noise) or 
whether signatures of nonlinear correlations are present, one of the most
direct approaches is based on the idea of surrogate data sets introduced by 
Theiler et al. (1992; see Kantz, \& Schreiber 1997 for a review). In simple words, this technique can be summarized 
as follows:\\ 
1) Assume as null 
hypothesis that the original time series is linear.\\
2) Construct
{\it linear} surrogate data that have the same linear characteristics of
the original data. In other words, the PSD of surrogate data should be 
indistinguishable from that of the original data, since linear processes are 
by definition completely characterized by the PSD or alternatively by the
autocorrelation function.\\
3) Use an appropriate nonlinear statistic to test the null hypothesis
comparing original and surrogate data. If this
test yields consistent values for surrogates and real data, we conclude that 
the original time series is linear in nature. On the other hand, if the 
value of the nonlinear statistic for the real data is significantly different 
from the corresponding values obtained using surrogates, then we infer the 
presence of nonlinearity.

Before presenting the results from this test, it is instructive to describe
a few basic details of the technique used to create surrogate data and the
main characteristics of the nonlinear statistic used in this case.

The most popular algorithms used to generate an ensemble of surrogate 
realizations are the Amplitude Adjusted Fourier Transform (AAFT) 
and the Iterative Amplitude Adjusted Fourier Transform (IAAFT) 
algorithms (Theiler et al. 1992; Schreiber \& Schmitz 1996). Although 
the AAFT and IAAFT algorithms conserve the amplitude distribution in 
real space and reproduce the PSD of the original data set quite accurately, 
it has been shown recently that both algorithms  may induce 
unwanted correlations in the Fourier phases (R\"ath \& Monetti 2008).
To guarantee that the surrogates in this study
are free from any higher order correlations,
we generate them  in the following way:
First, the time series is mapped onto a Gaussian distribution in a 
rank-ordered way, which means that the amplitude distribution of the 
original 
times series in real space is replaced by a Gaussian distribution in such a 
way that the rank-ordering is preserved, i.e. the lowest value of the 
original distribution
is replaced with the lowest value of the Gaussian distribution etc.
By applying this remapping we automatically focus on the temporal correlations 
of the data, while excluding any contributions to nonlinear correlations 
stemming from
the non-Gaussianity of the original intensity distribution.
Second, we Fourier transform the remapped time series, replace
the original phases by a new set of uniformly distributed and uncorrelated 
phases and perform an inverse Fourier transformation.
Note that the surrogate time series generated in such a way preserve 
{\it exactly} the power 
spectrum, while explicitly controlling the randomness of the phases.
For each time series under study we generate $50$ corresponding surrogates.
\begin{figure}
\begin{center}
\includegraphics[angle=0,width=9.cm]{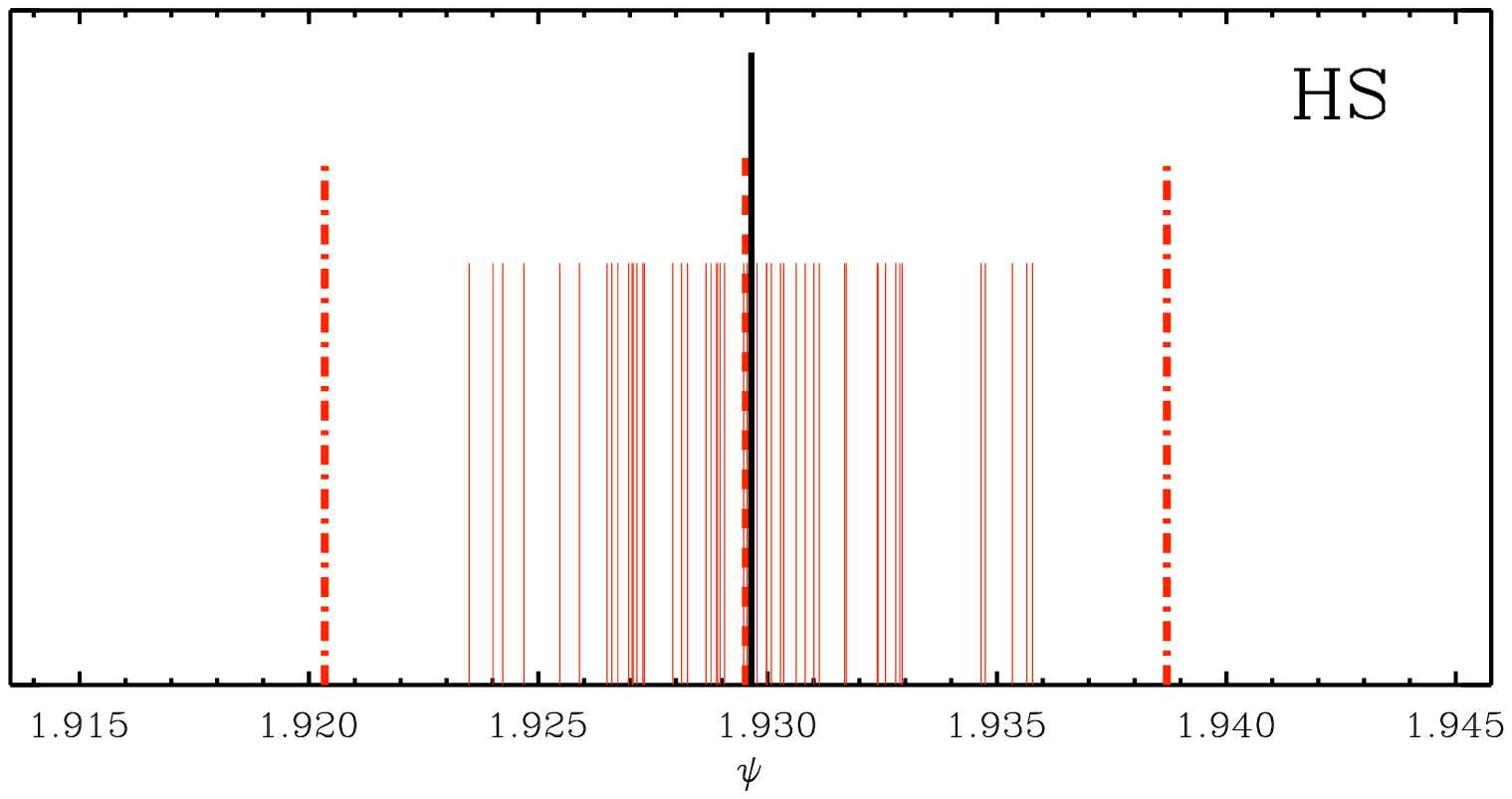}
\includegraphics[angle=0,width=9.cm]{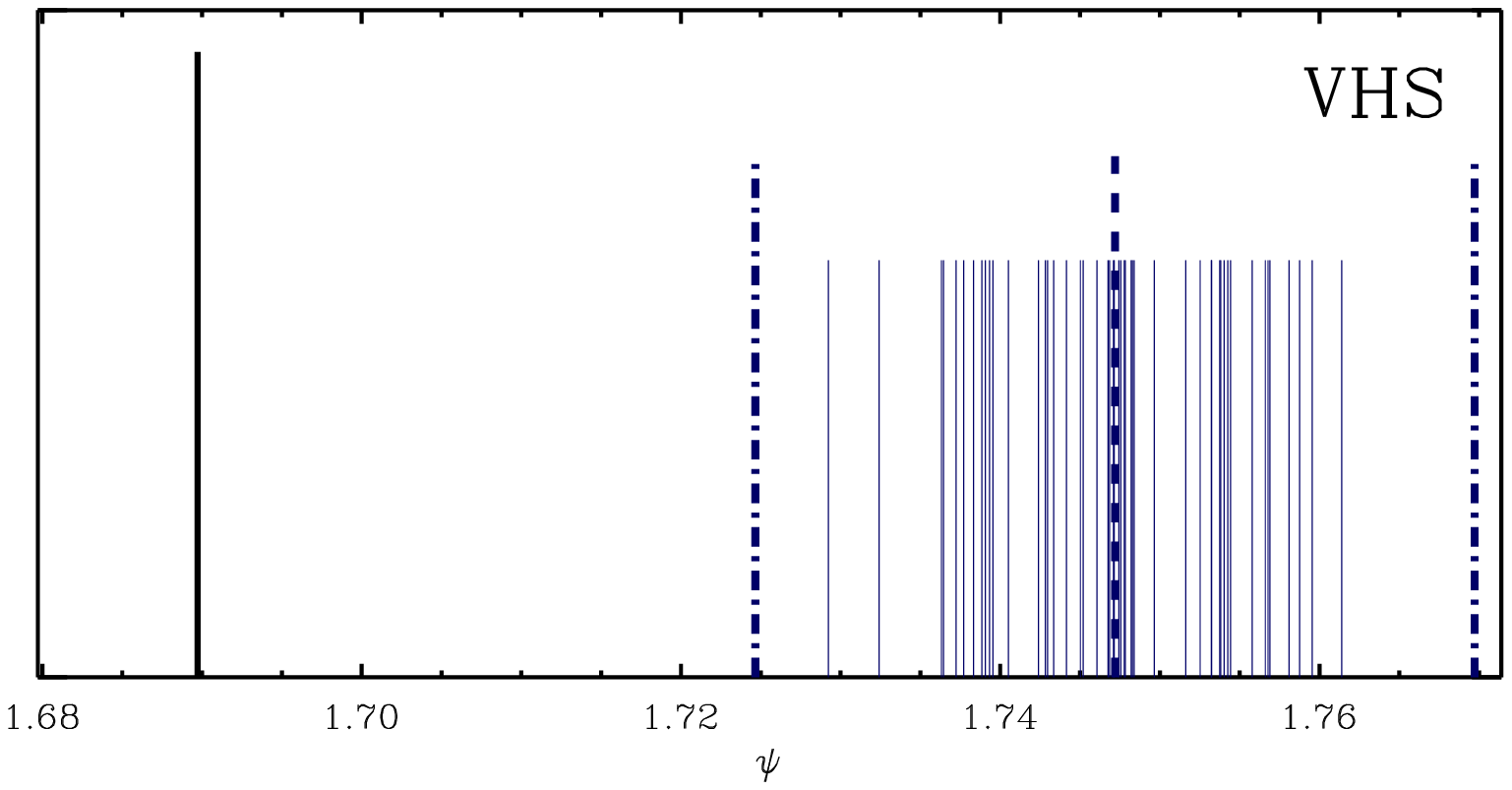}
\includegraphics[angle=0,width=9.cm]{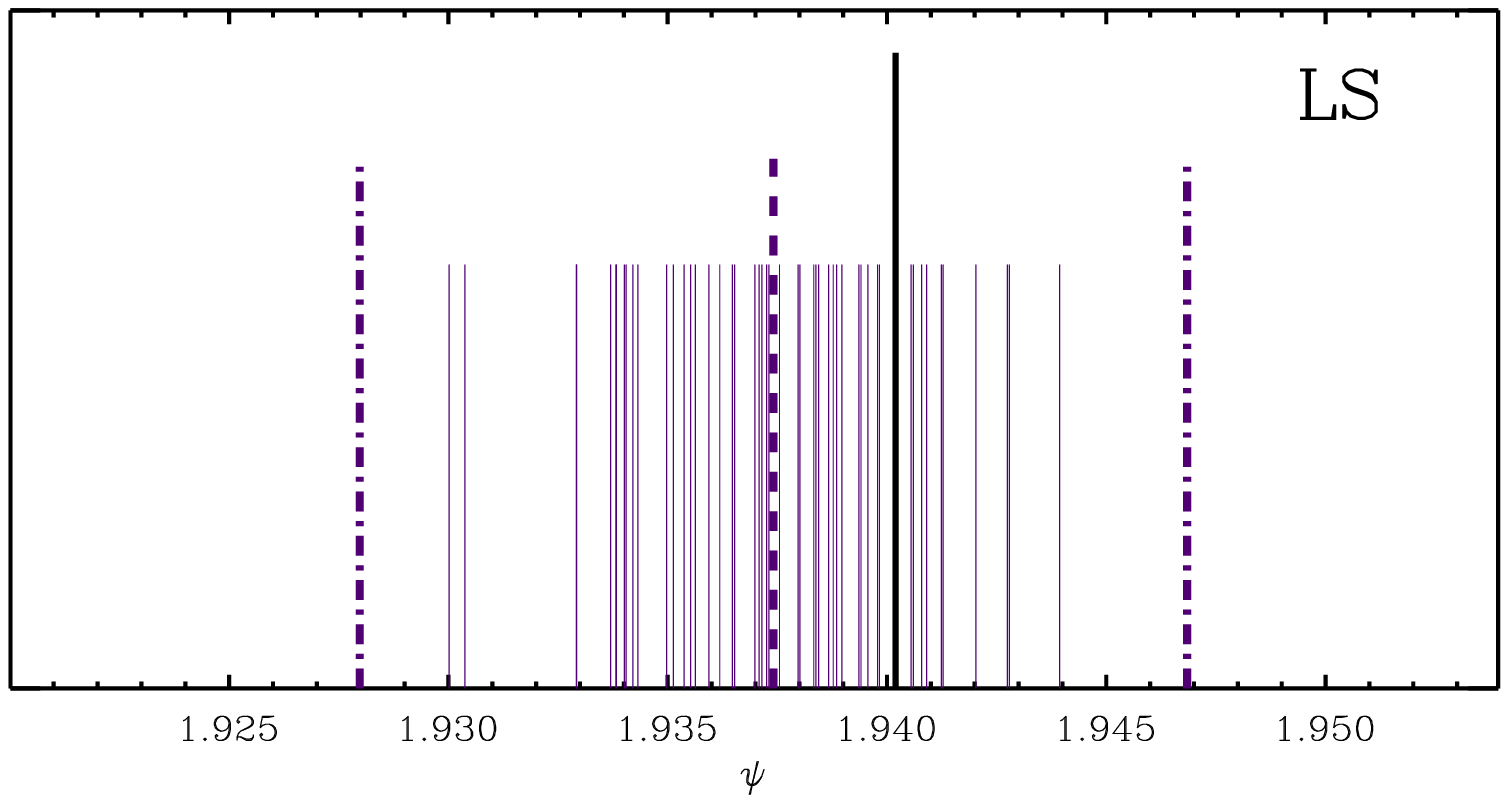}
\caption{Nonlinear prediction error (NLPE) for the  HS (top panel), VHS (middle 
panel), and LS (bottom panel). The black lines show the NLPE for the original time series.
The smaller thin colored lines denote $\psi$ for the respective surrogates. The thicker and longer
dashed line indicate the mean value $\langle \psi \rangle$  as derived from the 50 surrogate 
realizations.  The dash-dotted lines mark the $3 \sigma_{\psi}$ interval, i.e.  $\langle \psi \rangle - 3  \sigma_{\psi}$ and
$\langle \psi \rangle + 3  \sigma_{\psi}$, where the standard deviation $\sigma_{\psi}$ is also derived from the surrogates.}
\end{center}
\label{figure:fig8}
\end{figure}

Although in principle any nonlinear statistics may be used to compare real
and surrogate data, a systematic study performed by Schreiber \& Schmitz
(1997) indicates that the nonlinear prediction error (NLPE) is one of the most effective indicators of nonlinearity
and has good discrimination power to detect any deviation from
a Gaussian linear stochastic process. Therefore,
to test the presence of nonlinearity in \4u\ light curves we make use 
of the NLPE.

The equation used to compute the NLPE (hereafter $\psi$) as well as some
technical details are provided in Appendix B. Here we simply describe in a 
qualitative way the main characteristics of this nonlinear indicator.
Since this method relies on the time delay embedding technique described in 
$\S3$, it can exploit the direct correspondence between the scalars of
the starting 
time series, $x_1,~,x_2,~...,x_n$, and the corresponding set of vectors 
in the pseudo phase space:
${\bf x_1,~,x_2,~...,x_n}$. Specifically, to predict the ``future'' 
measurement
$x_{n+10}$ (i.e., the value of the time series 10 steps ahead of $x_{n}$), one 
must find the closest vector to ${\bf x_n}$, which we will
call ${\bf x_i}$ and corresponds to the scalar $x_i$, and then use the 
scalar $x_{i+10}$ as a predictor for
$x_{n+10}$. As explained in Appendix B, the rigorous process is slightly more 
complicated and involves several vectors in the neighborhood of ${\bf x_n}$, 
which in turn will yield several predictors.
The nonlinear prediction error $\psi$ is then provided by the average of
the differences between the actual value $x_{n+10}$ and the different 
predictor 
values.

We have carried out this test for all the light curves relative to the 2002
outburst of \4u, although for clarity reasons in Figure 8 we only show the 
results for the 3 representative light curves of the HS, VHS, and LS 
introduced 
in $\S$3. From this figure, it is clear that the absolute values of $\psi$
for the HS and IS are considerably larger than those measured in the VHS.  
This is an expected behavior, because in the VHS the time series is much more 
correlated with linear components showing up as QPO. As a consequence,
its predictability increases leading to lower values of $\psi$. More 
importantly, Figure 8 reveals that the VHS shows highly significant 
signatures for nonlinear correlations, unlike the HS and IS for which the 
values of $\psi$ obtained using real data are fully consistent with the 
values derived using linear surrogates.

Analyzing all time series belonging to the different spectral states we infer 
that all the HS and LS light curves are linear, whereas  
the 2 VHS light curves show respectively highly significant 
and marginally significant signs of nonlinearity. Specifically, the time 
series on MJD=52461, which corresponds to the
absolute minimum of the scaling index, shows the strongest and the only
statistically significant evidence ($\sim 7.5 \sigma$) for nonlinearity, as 
illustrated by the middle panel of Figure 8. A day before, on MJD=52460, the
evidence of nonlinearity is marginal (at a $\sim 2.4 \sigma$ significance 
level). Finally, the light curve on MJD=52459 that caught \4u\ during the 
HS-to-VHS transition (see Fig. 7) shows no indications of nonlinearity at all.

In conclusion, the surrogate test reveals that: 1) All HS and LS light 
curves are linear, 2) non-linearity 
indications appear during the VHS light curves, and the highest and most 
significant signal for non-linearity occurs for this light curves with the 
strongest QPO and the lowest value of $\langle\alpha\rangle$.
On one hand, the latter result suggests 
that the physical mechanism leading to strong QPOs is intrinsically
nonlinear, implicitly disfavoring QPO linear models. On the other hand, 
it indicates that the low
value of $\langle\alpha\rangle$ measured in the VHS is at least partially 
related to the presence of nonlinear correlations, which cannot be
detected by linear timing techniques.

\section{Summary and Conclusions}
We have carried out a nonlinear analysis of the variability properties
of the X-ray nova GBH \4u.
The main results can be summarized as follows:

\begin{itemize}
 
\item  We have used the WSIM to assign a single number, the mean scaling 
index $\langle\alpha\rangle$, to each individual light curve of the source. 
The large number of data when the source was in its HS, showed that the 
resulting $\langle\alpha\rangle$ values remain roughly constant, irrespective
of large changes in flux associated to the disk and PL components.

\item Similarly, the mean scaling index values remained roughly constant
during the VHS and LS, showing the following relationship:  
$\langle\alpha\rangle_{\rm VHS}~ < ~\langle\alpha\rangle_{\rm LS}~ <
\langle\alpha\rangle_{\rm HS}$. These results, which need to be confirmed 
for other GBHs, suggest that the mean scaling index $\langle\alpha\rangle$
may be used to parametrize the timing properties of an accreting source, 
and that it may be a true indicator of ``state'' in these systems.

\item When plotted versus time, the $\langle\alpha\rangle$ trend shows
no direct correlation with the total flux temporal behavior, which is
dominated by the accretion disk emission. On the other hand, 
the temporal evolution of $\langle\alpha\rangle$ appears to be
somewhat related to that of the power-law spectral
component: $\langle\alpha\rangle$ reaches its absolute minimum roughly 
at the same time as the PL flux reaches its absolute maximum, which occurs
during the VHS state. 

\item The search for nonlinearity using surrogate data and NLPE reveals that
the variability is linear in all light curves with the notable exception
of one observation in the VHS, which corresponds to the absolute minimum 
of $\langle\alpha\rangle$ and is also characterized by the presence of a 
strong QPO. 
\end{itemize}

An important implication of the detection of nonlinearity in the VHS 
is that all intrinsically
linear models proposed to produced low frequency QPOs (LFQPOs)  
may be ruled out for  \4u. 
The formation of QPOs (at both high and low frequencies) in GBHs is 
currently a matter of debate, as several viable competing models have been 
proposed (see, e.g.,  McClintock \& 
Remillard  2006 for a review). It is therefore
important to derive model-independent constraints that may break or at least
reduce the current model degeneracy. 
If the detection of nonlinearity associated
with strong LFQPOs is confirmed in other GBHs, then intrinsically linear
models such as the global disk oscillations proposed by Titarchuk \&
Osherovich (2000) can be safely ruled out, whereas alternative models such as
the shock oscillation model (Chakrabarti \& Manickam 2000) or the 
accretion-ejection instability model (Tagger \& Pellat 1999), which
may account for nonlinearities, are still viable solutions.

In summary, the findings derived from this work suggest that this kind of
 nonlinear analysis can be useful in the field of GBHs and complement
the temporal analysis carried out with linear techniques. Specifically, 
the simplicity of the WSIM,
which characterizes the global variability properties of a light curve via
a single number,
$\langle\alpha\rangle$, suggests that this technique might be 
successfully applied to study
correlated temporal and spectral variability. In addition, the 
robustness of the WSIM, that performs well also with noisy data and 
relatively short light curves, naturally suggests an application to AGN 
variability with the possibility of direct comparison with GBHs. 

Since we have limited our analysis to \4u\ only, before deriving
any general conclusions, we should carry out a systematic nonlinear analysis 
on several GBHs during their spectral transition.
Specifically, it will be important to assess whether
specific values of $\langle\alpha\rangle$ are associated with specific 
spectral states (e.g., 
$\langle\alpha\rangle_{\rm VHS}\simeq 1.8 $,
$\langle\alpha\rangle_{\rm HS}\simeq 2.0$,
$\langle\alpha\rangle_{\rm HS}\simeq 2.05 $), or if only the temporal
trend of $\langle\alpha\rangle$ during the outburst (i.e., the presence of
a pronounced minimum during the VHS) is similar to the one displayed by \4u. 
To this aim we will carry out a similar analysis on a few prominent
GBHs with outbursts that are completely covered by the \rxte\ PCA and where 
the different spectral state are well sampled. 
This test will unequivocally reveal whether 
$\langle\alpha\rangle$ can be used as reliable indicator of spectral states.
In addition, the planned study will provide useful information on
correlated variability of $\langle\alpha\rangle$ and several relevant spectral
parameters, and on the presence of nonlinearity in different spectral states.

{}

\appendix
\section{Dependence of SIM on $\tau$, $D$, and $R$}
As explained in $\S3$, the WSIM depends on 3 parameters: 
time delay $\tau$ and embedding dimension $D$, which are related to the 
phase space reconstruction process, and the radius
$R$ at which the logarithmic derivative (i.e., the scaling index) is computed.

\begin{figure}
\includegraphics[bb=85 55 445 545,clip=,angle=0,width=8.cm]{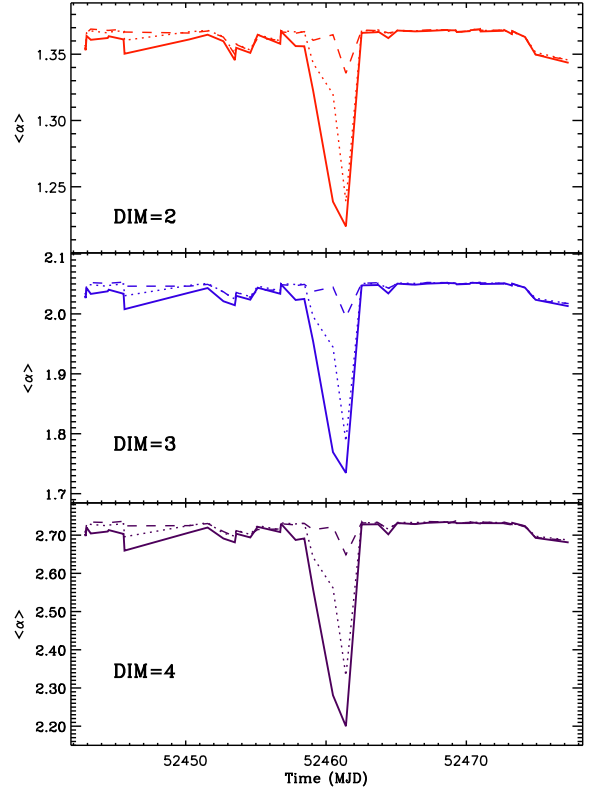}
\caption{Temporal evolution of the mean WSIM during the 2002 flare of 
\4u\ for embedding dimensions 2 (top panel), 3
(middle panel), and 4 (bottom panel). Solid, dotted, and dashed lines refer 
to $\tau=0.1,~1$, and 10 s, respectively.
}
\label{figure:figA1}
\end{figure}

In this section, we visually demonstrate that the impact of these parameters
on our main findings is basically irrelevant for a broad range of reasonable
choices. Figure.~\ref{figure:figA1} shows the temporal evolution of
$\langle\alpha\rangle$ for  $D=$2 (top panel), 3 (middle panel), and 4 (bottom
panel). Solid lines refer to $\tau=0.1$ s, dotted lines to $\tau=1$ s, and 
dashed lines to $\tau=100$ s; in this case the radius is fixed at 1.6. 
It is noticeable that all the values of 
$\langle\alpha\rangle$ consistently increase as the embedding dimension
increases. This reflects the fact that a significant part of the variability
is random and this random component translates into larger values of $\alpha$
as the dimension of the embedding space increases.

From  Fig.~\ref{figure:figA1} it is evident that
all different combinations of $\tau$ and $D$
reproduce the same temporal evolution trend of
$\langle\alpha\rangle$ with a large central dip preceded by to low-amplitude
dips and followed by a final steady decrease. It is also clear that using
long time delays, such as $\tau=100$ s, the dips appear less prominent. This
simply reflects the fact that long time delays necessarily loose the 
information relative to short-term temporal correlations.

\begin{figure}
\includegraphics[bb=35 30 335 350,clip=,angle=0,width=8.cm]{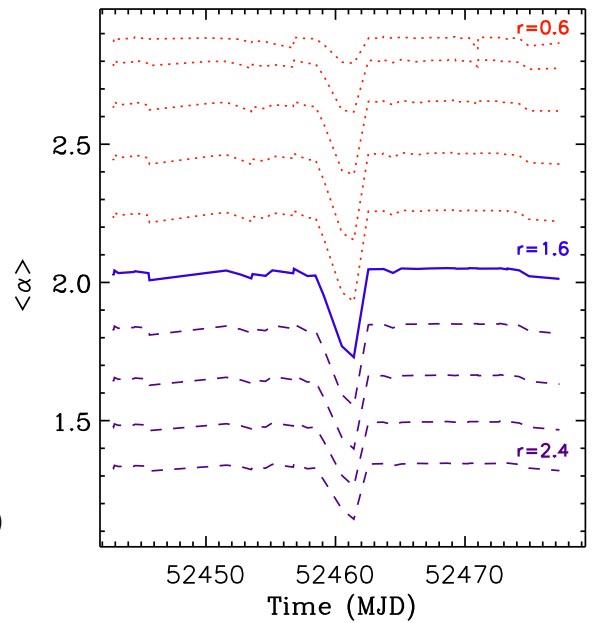}
\caption{Temporal evolution of the mean WSIM during the 2002 flare of 
\4u\ for embedding dimensions 3, with $r$ ranging from 0.6  (top dotted line)
to 2.4 (bottom dashed line).
}
\label{figure:figA2}
\end{figure}

Figure.~\ref{figure:figA2} illustrates the impact of the radius $R$ on the
temporal evolution of $\langle\alpha\rangle$ (this time $D=3$ and 
$\tau=0.1$ s). Once again, all values of $R$ are able to recover the 
same temporal evolution trend of $\langle\alpha\rangle$ described above.


\section{Nonlinear Predictor Error}
To calculate NLPE, the time series is embedded in a $D$-dimensional space 
using the method of delay coordinates as described in section 3. We use here
also the embedding dimension $D=3$. The delay time $\tau$ was determined using
the criterion of zero crossing of the autocorrelation function considering 
the VHS, where the ACF is sufficiently different from a random process.
The NLPE is defined as
\begin{equation}
\psi = \frac{1}{M-T-(D-1)\tau} \left( \sum_{n=(D-1)\tau}^{M-1-T} [\vec{x}_{n+T} -F(\vec{x}_n)]^2 \right)^{1/2} 
\end{equation}
where $F$ is a locally constant predictor (i.e., a quantity that 
remains constant for a local surrounding of a point under study in the 
D-dimensional embedding space), $M$ is the length of the time series, 
and $T$ is the lead time (i.e., the number of time steps ahead of
the considered one for which we want to make a prediction).
The predictor $F$ is calculated by averaging over future values 
of the $N \, (N=D+1)$ nearest neighbors in the delay coordinate 
representation. We have studied the behavior of $\psi$ as a function 
of the lead time $T$ and did not find significant variations. We set this 
value to  $T=10$ time steps. The results of this analysis for the 3 
representative light curves of the HS, VHS, and LS are shown in
Figure 8.
\end{document}